\documentclass{elsart}
\usepackage{amsmath}
\usepackage{amssymb}
\usepackage{amsmath}
\usepackage{verbatim}
\usepackage{epsfig}

\newcommand{\new}{\newcommand} 
 \new{\np}{\newpage} 
 \new{\md}{\medskip} 
 \new{\sm}{\smallskip}
\new{\ra}{\rightarrow} 
\new{\la}{\leftarrow}
\new{\Ra}{\Rightarrow} 
\new{\La}{\Leftarrow}
 \new{\cen}{\center}

\new{\eg}{for example}
\new{\Txs}{There exists}
\new{\txs}{there exists}
\new{\tx}{there exist} 
\new{\Tx}{There exist}
\new{\ter}{interpretation}
\new{\ters}{interpretations}
\new{\st}{such that}
 \new{\ifa}{if and only if} 
 \new{\ass}{assignment} 
 \new{\asss}{assignments}
\new{\equ}{equivalent} 
\new{\Wma}{We may assume} 
\new{\wma}{we may assume}
\new{\wrt}{with respect to}
\new{\enum}{\begin{enumerate}}
\new{\enumE}{\end{enumerate}}
\new{\sat}{satisfy}
\new{\sating}{satisfying}
\new{\otoh}{on the other hand}
\new{\Otoh}{On the other hand}
\new{\ih}{induction hypothesis}
\new{\iha}{induction hypothesis applied}

\new{\be}{\begin} 
\new{\proofend}{$\;\square$ \bigskip}
\new{\proof}{\noindent{\it Proof. }}

\new{\as}{\,{\tt {:=}}\,}
\new{\whi}{\,\mathit{while}\,}
\new{\du}{\,\mathit{do}\,}
\new{\then}{\,\mathit{then}\,}
\new{\els}{\,\mathit{else}\,}
\new{\beginn}{\text{\it begin }}
\new{\eend}{\text{\it end }}
\new{\tru}{{\mathsf T}}
\new{\fal}{{\mathsf F}}
\new{\si}{\,\mathit{if}\,}
\new{\ski}{\mathit{skip}}
\new{\sch}{\mathit{Sch}}  
\new{\Int}{\mathit{Int}}
\new{\term}{\mathit{Term}} 
\new{\sig}{\mathit{Ass}}

\new{\p}{\mathcal{P}}
\new{\f}{\mathcal{F}}
\new{\labels}{{\mathcal L}}

 \new{\var}{\mathcal{V}} 
\new{\emp}{\varepsilon}

\new{\tl}{\triangleleft}
 \new{\fA}{\Gamma}
 \new{\fS}{S}
 \new{\fT}{T}
 \new{\lang}{{\mathit alphabet}}
 \new{\allpath}{\Pi^\omega}
 
 \new{\pathset}{\Pi}
 \new{\pathinf}[1]{\Pi^{\infty}(#1)} 
 \new{\path}[2]{{\pi}_{#1} ( #2)}        
 
 \new{\set}[1]{\{ #1\}}
 
 \new{\subterms}{\mathit{Subterms}}
\new{\terms}{\mathit{Terms}}
 \new{\termfun}{\mathit{TermSymbols}}
 \new{\pathlets}{\mathit{PathSymbols}}
 \new{\predpass}{\mathit{Passpred}}
 \new{\funpass}{\mathit{Passfunc}}
 
 \new{\terlets}[1]{{\mathit{InterLetters}}_{#1}}
 \new{\sub}[1]{\,{\mathit{sub}}_{#1} \,}
 \new{\subtrans}[1]{\,{\mathit{sub}^*}(#1) \,}
 
 \new{\state}{{\rm State}}
 \newcommand{\quine}[1]{ [ \!  [{#1}] \!  ] }
 
 \new{\arity}{\mathit{arity}}
 
 \new{\schema}{\mathit{schema}}
 
 \new{\predpath}[1]{\mathit{Predpaths}_{\, #1}}
\new{\func}{\mathit{Funcs}}
 \new{\pred}{\mathit{Preds}}
 \new{\ifpred}{\mathit{ifPreds}}
 \new{\whipred}{\mathit{whilePreds}}
 \new{\dd}[2]{{\mathcal M}\quine{#1}^{#2}_d}
 \new{\ee}[2]{{\mathcal M}\quine{#1}^{#2}_e}     
  \new{\mean}[3]{{\mathcal M}{\quine{#1}^{#2}_{{#3}}}}  
 \new{\supe}[3]{#1_{#2}^{#3}} 
 \new{\pre}{\mathit{pre}}
\new{\maxpre}{\mathit{maxpre}}
 
\newcommand\vx{{\mathbf{x}}}
\newcommand\vy{{\mathbf{y}}}
\newcommand\vz{{\mathbf{z}}}
\new{\vt}{{\mathbf{t}}}

\newcommand{\vu}{\mathbf{u}}

\newcommand{\vv}{\mathbf{v}}
\newcommand{\vw}{\mathbf{w}}

\new{\ul}{\underline}

\new{\Perc}[1][S]{ !!\rightsquigarrow_{#1} }
\new{\finperc}[1][S]{\rightsquigarrow_{#1}^{\text{final}}}

\new{\cont}[1][S]{ \, \searrow_{#1} \, }
\new{\tcont}[1][S]{ \, {\searrow_{#1}}^{*} \; }

\new{\thru}[1][\fS]{\longmapsto_{#1}}
\new{\weak}[1]{\cong_{#1}}
\new{\strong}[1]{\mathbf{\cong_{#1}}}
\new{\seq}{\mathit{Sequences}}
\new{\mseq}{\mathit{maxSequences}}
\new{\isdefd}{\stackrel{\mathrm{def}}{=}}
\new{\assig}[1][S]{ \mathit{assign}_{#1} }
\new{\refset}[1][S]{ \mathit{refVars}_{#1} }
\new{\refvec}[1][S]{ \mathbf{refvec}_{#1} }

\new{\sli}[1][V]{ {{\mathit{sl}_{#1}}} }
\new{\inv}[1][S]{\mathit{Inv}_{#1}}
\new{\comps}{\mathit{Comps}}
\new{\opn}{ < \! \!\!}
\new{\clo}{\!\! >}
\new{\eqspace}{\!\! =\!\!}
\new{\simi}[1][R,V]{\tilde_{#1}}
\new{\need}[1][S]{{\mathcal N}_{{#1}}}
\new{\tword}[3]{\opn{#1}\supe{({#2})}{}{#3}\clo}
\new{\occ}{\mathit{Occ}}
\new{\gr}[1][S]{\mathit{grade}_{#1}}
\new{\grx}[1][S]{\mathit{grade}^{\mathit EXT}_{#1}}
\new{\grint}[1][S]{\mathit{grade}^{\mathit INT}_{#1}}
\new{\dex}[1][S]{\mathit{index}_{#1}}
\new{\dexx}[1][S]{\mathit{index}^{\mathit EXT}_{#1}}
\new{\inte}{\mathit{int}}
\new{\exte}{\mathit{EXT}}
\new{\seg}{\mathit{Seg}}
\new{\comb}{\bowtie}
\new{\letter}[2]{\opn {#1}\,{\clo}\!\!_{#2}\,}
\new{\pletter}[2]{\opn {#1},#2\,{\clo}}
\new{\body}[1][S]{ {{\mathit{body}_{#1}}} }
\new{\tf}{\{\tru,\fal\}}
\new{\op}{\neg}

\new{\outif}[1][S]{ {{\mathit{outif}_{#1}}} }
\new{\outwhi}[1][S]{ {{\mathit{outwhile}_{#1}}} }
\new{\out}[1][S]{ {{\mathit{out}_{#1}}} }
\new{\throu}[1][S]{ {{\mathit{thru}_{#1}}} }
\new{\back}[1][S]{ {{\mathit{back}_{#1}}} }
\new{\con}{\mathcal C}
\new{\varr}{\mathit whileVar}
\new{\corr}[1][S,T]{\mathit{Corr}_{#1}}
\new{\scorr}[1][S,T]{\mathit{StrongCorr}_{#1}}
\new{\precc}[1][S]{<\!\!<_{#1}}
\new{\abuv}[1][S]{\mathit{above}_{#1}}

\new{\tail}[1][S]{\mathit{tail}_{#1}}
\new{\head}[1][S]{\mathit{head}_{#1}}
\new{\tailterm}[1][S]{\mathit{Tailterms}_{#1}}
\new{\headterm}[1][S]{\mathit{Headterms}_{#1}}
\new{\com}[1][S]{\mathit{tail}_{#1}}

\new{\simbol}{\mathit{symbol}}
\new{\lsimbol}{\mathit{symbol}^{(\l)}}
\new{\sym}[1][S]{\mathit{Symbols}_{#1}}
\new{\lsym}[1][S]{\mathit{{Symbols}^{\mathcal L}}_{#1}}
\new{\ifprec}[1][p,S]{\sqsubseteq_{#1}}
\new{\pass}[1][S]{\mathit{passage}_{#1}}
\new{\Pass}{\mathit{Passages}}
\new{\init}{\mathit init}
\new{\next}[1][S]{{{\mathit{nextsymbol}}_{#1}}}

\new{\fir}{\mathit{firstpred}}
\new{\swhi}{\mathit{swhilePreds}}
\new{\ia}{immediately above}
\new{\sw}{super-while}

\new{\simil}[1][u]{\,\mathord{\mathit{simil}_{#1}\,}}
\new{\class}{\mathit{class}}
 \new{\trunc}[1][S]{\mathit{trunc}_{#1}\,}
 \new{\dep}[1][S]{\mathit{depnum}_{#1}\,}

 \new{\px}{p(\vx)}
 \new{\py}{p(\vy)}
 \new{\qx}{q(\vx)}
 \new{\qy}{q(\vy)}

 \new{\subb}{\subseteq}
  \new{\supp}{\supseteq}

  \new{\goto}{\mathit{goto\,\,}}

 \new{\ppair}[1][S]{\mathit{Predpairs}_{#1}}

\new{\cng}[1][u]{\mathit{cong}_{#1}\,}

\new{\simclass}[2][S]{[#2]_{#1}}

\new{\sto}{\mathit{STOP}}
\new{\start}{\mathit{START}}

\new{\before}[1][S]{<\!\!<_#1}

\new{\weis}[1][S]{{\mathcal W}_{#1}}
\new{\wpath}[1][S]{{\mathcal W}{\mathit paths}_{#1}}
\new{\wpred}[1][S]{{\mathcal W}{\mathit preds}_{#1}}
\new{\wfunc}[1][S]{{\mathcal W}{\mathit funcs}_{#1}}
\new{\wsym}[1][S]{{\mathcal W}{\mathit symbols}_{#1}}

\new{\lfnl}{LFn-L}
\new{\lib}{{\mathit libFuncs}}
\new{\cfunc}{{\mathit constFuncs}}

\new{\ls}[2][l]{{#2}^{(#1)}}
\new{\labs}{\mathit{labels}}          

 \new{\lfunc}{{\func^{\mathcal L}}}
\new{\lpred}{\pred^{\mathcal L}}
 \new{\lifpred}{{\ifpred^{\mathcal L}}}
 \new{\lwhipred}{{\whipred^{\mathcal L}}}

\new{\resp}{respectively}

\new{\gra}{\mathit Graph}
\new{\simcl}[2][S]{[#2]_{#1}}

 \new{\last}{\mathit{end}}
\new{\proj}[1][S']{ {\mathit{proj}}_{#1} }

\new{\scheq}{\text{:}\!\!\!=}

\new{\ubef}{\before[\text{}]}
\new{\usclass}[1]{[#1]}

\new{\call}{{\mathit call}}
\new{\uncall}{{\mathit uncall}}

\new{\conc}{{\parallel}}

\new{\sed}[1][S]{ \exists {\mathit{DD}}_{#1}    }

\new{\sad}[1][S]{ \forall {\mathit{DD}}_{#1}    }

\new{\strt}{\mathbf{start}}
\new{\nd}{\mathbf{end}}

\new{\emptwrd}{\varepsilon}

\new{\frst}{\mathit{firstlabel}}

\new{\datdep}[1][S]{\exists\mathit{DatDep}_{#1}}
\new{\datdap}[1][S]{\forall\mathit{DatDep}_{#1}}
\new{\vrtces}{\mathit{Vertices}}

\new{\lop}{\mathbf{loop\;}}

\new{\reg}[1][{\Sigma}]{{\mathbf{Reg}} _#1           }
\new{\Reg}[1][{\Sigma}]{{\mathbf{Reg}} _#1 ^{\omega} }

\begin{document}




\begin{frontmatter}


\title{On the computational complexity of dynamic  slicing problems for program schemas}


\author[shef]{Michael R. Laurence}
\author[gold]{Sebastian Danicic}
\author[bru]{Robert M. Hierons}

\address[gold]{Department of Computing, Goldsmiths College,
University of London,
New Cross,
London SE14 6NW
UK}

\address[bru]{
Department of Information Systems and Computing,
 Brunel University,
 Uxbridge, Middlesex, UB8 3PH. }

\address[shef]{ Department of Computer Science,
Regent Court, 211 Portobello, Sheffield, S1 4DP, UK.}



\maketitle

 \begin{abstract}
\noindent
Given a program, a quotient can be obtained from it by deleting zero or more statements. The field of program slicing is concerned with  computing a quotient of a program which preserves part of the behaviour of the original program.  All program slicing algorithms take account of the structural properties of a program such as control dependence and data dependence rather than the semantics of its functions and predicates, and thus  work, in effect,  with program schemas. The  dynamic slicing criterion of Korel and Laski requires only that program behaviour is preserved in cases where the original program follows a particular path, and that the slice/quotient follows this path.  In this paper we formalise Korel and Laski's definition of a dynamic slice as applied to linear schemas, and also formulate a less restrictive definition in which the path through the original program need not be preserved by the slice. The less restrictive definition has the benefit of leading to smaller slices. For both definitions, we compute complexity bounds for the problems of establishing  whether a given slice of a linear schema is a dynamic slice and whether a linear schema has a non-trivial dynamic slice and prove that the latter problem is NP-hard in both cases.  We also give an example to prove that minimal dynamic slices (whether or not they preserve the original path) need not be unique.
 \end{abstract}

\begin{keyword}
program schemas
\sep
program slicing
\sep
NP-completeness
\sep
Herbrand domain
\sep
linear schemas
\end{keyword}

\end{frontmatter}



\section{Introduction}   \label{intro.sect}

A schema represents the statement structure of a program by replacing real functions and predicates by symbols representing them.
A schema, ${S}$, thus defines a whole class  of programs which  all have the same
structure. A schema is \emph{linear} if it does not contain more than one occurrence of the same function or predicate symbol.
As an example, Figure \ref{S.fig} gives a schema $S$; and  Figure \ref{program.fig} shows one of the programs obtainable from the schema of Figure \ref{S.fig} by interpreting its function and predicate symbols.
\\
The subject of schema theory  is connected with that of program  transformation and was originally motivated by the wish to compile programs     effectively\cite{greibach:theory}.
 Thus an   important problem in schema theory is that of establishing whether two schemas are equivalent; that is, whether they always have the same termination behaviour, and give the same final value for every variable, given any initial state and  any  \ter\ of   function and predicate symbols. In Section  \ref{schema.equiv.sect}, the history of this problem is discussed.

\begin{figure}[t]
\begin{center}
$
\begin{array}{llll}
u\as h(); \\
\si  p(w) && \then
& v\as f(u); \\
&& \els
&  v\as g();
 \end{array}
 $
\end{center}
\caption{A schema} \label{S.fig}
\end{figure}

\bigskip

\begin{figure}[b]

\begin{center}
$
\begin{array}{llll}
& \\
u\as 1; \\
\si  w> 1 && \then
& v\as u+1; \\
&& \els
&  v\as 2;
 \end{array}
 $
\end{center}
\caption{A program  defined from the schema  of Figure \ref{S.fig}} \label{program.fig}
\end{figure}

 Schema theory is also  relevant to program slicing, and this is the motivation for the main results of this paper.
We define a \emph{quotient}  of a schema $S$  to be any schema obtained   by deleting zero or more statements from $S$. A quotient of $S$ is non-trivial if it is distinct from $S$.
Thus a quotient of a schema is not required to satisfy any semantic condition; it is defined purely syntactically. The field of program slicing is concerned with  computing a quotient of a program which preserves part of the behaviour of the original program. Program slicing is used in program comprehension \cite{delucia:understanding,mhetal:icsm01}, software maintenance \cite{canfora:salvaging,cimitile:specification,gallagher:surgeon,gallagher:maintenance}, and debugging \cite{agrawal:debugging,kamkar:thesis,lyle:debugging,weiser:experiments}.

All program slicing algorithms take account of  the structural properties of a program such as control dependence and data dependence rather than the semantics of its functions and predicates, and
thus  work, in effect,  with linear  program schemas. There are two main forms of program slicing; static and dynamic.
\begin{itemize}
\item
In static program slicing,  only the  program itself is used to construct a slice. Most static slicing algorithms are based on Weiser's algorithm\cite{weiser:slicing84}, which uses the data and control dependence relations of the program in order to  compute the set of statements which the slice  retains. An end-slice of a program \wrt\ a variable $v$ is a slice that always returns the same final value for $v$ as the original program, when executed from the same input.   It has been proved that  Weiser's algorithm gives minimal static end-slices\cite{sdetal:lpr} for linear, free, liberal program schemas. This result has recently been strengthened by allowing function-linear schemas, in which only predicate symbols are required to be non-repeating\cite{laurence:flfl}.
\item
In dynamic program slicing,  a path through the program is also used as input. Dynamic slices of programs may be smaller than static slices, since they are only required to preserve behaviour in cases where the original program follows a particular path.  As originally formulated by
Korel and Laski  \cite{korel:dynamic-slicing},   a dynamic  slice of a program $P$ is defined by three parameters besides $P$, namely a variable set $V$,  an initial input state $d$ and an integer $n$. The slice \wrt\ these parameters is  required to follow the same path as $P$ up to the $n$th statement (with statements not lying in the slice deleted from the path through the slice) and
 give the same value for each element of $V$  as $P$ after the $n$th statement  after execution from the initial state $d$. Many dynamic slicing algorithms have been written \cite{agrawal:90dynamic,beszedes:dynamic,gopal:dynamic-slicing,kamkar:interprocedural:dynamic,kamkar:application,korel:dynamic-slicing,korel:gotos,korel:ist-paper}. Most of these compute a slice using the data and control dependence relations along the given path through the original program. This  produces a correct slice, and uses polynomial time, but need not give a minimal or even   non-trivial slice even where one exists.
\end{itemize}

Our definition of a path-faithful dynamic slice (PFDS) for a linear schema $S$ comprises two parameters besides $S$, namely a path through $S$ and a variable set, but not an initial state. This definition is analogous to that of Korel and Laski, since the initial state included in their parameter set is  used solely in order to compute a path through the program in linear schema-based slicing algorithms. We prove, in effect, that it is decidable in polynomial time whether a particular quotient of a program is a dynamic  slice in the sense of Korel and Laski, and that the problem of establishing whether a program has a non-trivial path-faithful dynamic  slice is intractable, unless P$=$NP. This shows that there does not exist  a tractable dynamic  slicing  algorithm  that produces correct slices  and always  gives  a  non-trivial slice of a program   where one exists.

 The requirement of Korel and Laski that the path through the slice be path-faithful may seem unnecessarily strong. Therefore
 we define a more general dynamic slice (DS), in which the sequence of functions and predicates through which the path through the slice passes is a subsequence of that for the path through the original schema,
 but the path through the slice must still pass the same number of times through the program point at the end of the original path. For this less restrictive definition, we prove  that it is decidable in Co-NP time whether a particular slice of a program is a dynamic  slice, and the problem of  establishing whether a program has a non-trivial  dynamic  slice is NP-hard.

 We also give an example to  prove that unique minimal  dynamic slices (whether or not path-faithful)  of a linear schema $S$ do not always exist.

The results of this paper have several practical ramifications. First, we prove that the problem of deciding whether a linear schema has a non-trivial dynamic slice is computationally hard and clearly this result must also hold for programs. In addition, since this decision problem is computationally hard, the problem of producing minimal dynamic slices must also be computationally hard. Second, we define a new notion of a dynamic slice that places strictly weaker constraints on the slice than those traditionally used and thus can lead to smaller dynamic slices. In Section \ref{gen.dyn.defn.sect} we explain why these (smaller) dynamic slices can be appropriate, motivating this through a problem in program testing. Naturally, this weaker notion of a dynamic slice is also directly applicable to programs. Finally, we prove that minimal dynamic slices need not be unique and this has consequences when designing dynamic slicing algorithms since it tells us that algorithms that identify and then delete one statement at a time can lead to suboptimal dynamic slices.

It should be noted that much theoretical
 work on program slicing and program analysis, including that of M\"{u}ller-Olm's study of dependence analysis of parallel programs~\cite{mueller-olm:precise.parallel}, and on deciding validity of relations between variables at given program points \cite{mueller-olm.seidl:poly.prog.inv,mueller-olm.seidl:interp.prog.lin.alg} only considers programs in which branching is treated as non-deterministic, and is thus more `approximate' than our own in this respect, in that we take into account control dependence as part of the program structure.

\subsection{Different classes of  schemas}   \label{diff.classes.schemas.sub}

Many subclasses of schemas have been defined:

\begin{description}
\item[Structured schemas,] in which \emph{goto} commands are forbidden, and thus loops must be constructed using while statements. \emph{All schemas considered in this paper are structured.}
 \item[Linear schemas,]  in  which   each
function and predicate symbol occurs at most once.
\item[Free schemas,]  where all paths are executable under some interpretation.
\item[Conservative schemas,]
 in which every \ass\ is of the form  \\ $v\as f(v_1,\ldots, v_r);$ where $v\in \{v_1,\ldots, v_r\}$.
\item[Liberal schemas,] in which two \asss\ along any executable path can always  be made to assign distinct values to their respective variables by a suitable choice of domain.
\end{description}

It can be easily shown that all conservative schemas are liberal.

Paterson \cite{paterson:thesis}
gave a proof that it is decidable  whether a   schema is both liberal and  free;
 and since he also gave an algorithm transforming a schema $S$ into  a schema $T$ \st\ $T$ is both liberal and  free \ifa\ $S$ is  liberal, it is clearly decidable whether a schema is liberal. It is an open problem whether   freeness is decidable for the class of  linear   schemas. However he also proved, using a reduction from the Post Correspondence Problem,  that it is not decidable whether a schema is free.

\subsection{Previous results on the decidability of  schema equivalence} \label{schema.equiv.sect}

Most previous research on schemas has focused on schema equivalence.
All  results on the decidability of equivalence of schemas are either negative or confined to very restrictive classes of schemas. In particular
Paterson \cite{luckham:formalised} proved  that  equivalence
is undecidable for the class of all    schemas containing at least two variables,  using a reduction from the halting problem for Turing machines.
Ashcroft and Manna showed  \cite{ashcroft:while-goto-siam} that an arbitrary schema, which may include \emph{goto} commands, can be effectively transformed into an equivalent structured schema, provided that statements such as  $\whi \neg  p(\vu) \du T$ are permitted; hence Paterson's result shows that any class of schemas for which equivalence can be decided must not contain this class of   schemas.  Thus in order to get positive results on this problem, it is clearly necessary to define the relevant  classes of schema with great care.

Positive results on the decidability of equivalence of schemas include the following;
 in an early result in schema theory, Ianov \cite{ianov:logical} introduced a
restrictive class of schemas, the Ianov schemas,  for which equivalence
is decidable.  This problem was later shown to be NP-complete \cite{rutledge:ianov.schemata,constable:ianov.schemas}.
 Ianov schemas are characterised by being
monadic (that is, they  contain only a {\em single}
  variable) and having only unary function symbols; hence Ianov schemas are conservative.

 Paterson \cite{paterson:thesis} proved that equivalence is decidable
for a class of schemas called {\em progressive schemas}, in which every assignment
references the variable assigned by the previous assignment
along every legal path.

Sabelfeld \cite{sabelfeld:algorithm}
 proved that equivalence is
decidable for another class of schemas called {\em through
schemas}. A  through schema satisfies two conditions: firstly,
that on every path from an accessible predicate $p$ to a predicate $q$
which does not pass through another predicate,
and every variable $x$ referenced by $p$, there is a variable referenced by $q$ which defines a term containing the term defined by
 $x,$ and secondly,
 distinct variables referenced by a
 predicate can be made to  define distinct terms under some \ter.

 It has been proved that for the class of    schemas which are  linear, free  and conservative, equivalence is decidable \cite{mletal:cfl}.
More recently, the same conclusion was proved to hold under the weaker hypothesis of liberality  in place of conservatism  \cite{lfl:tr,mletal:lfl}.

\subsection{Organisation of the paper}

In Section \ref{basic.sect} we give  basic definitions of schemas. In Section  \ref{dyn.defn.sect} we define path-faithful dynamic  slices and in Section  \ref{gen.dyn.defn.sect} we define general dynamic  slices.  In Section  \ref{two.mins.sect} we give an example to prove that unique minimal  dynamic  slices need not exist. In Section  \ref{decide.dps.sect} we prove complexity bounds for problems concerning the existence of dynamic  slices. Lastly, in Section \ref{conclusion.sect}, we discuss further directions for research in this area.

\section{Basic Definitions of Schemas}  \label{basic.sect}

Throughout this paper,  $\f$, $\p$,  $\var$ and $\mathcal L$ denote
fixed infinite sets of \emph{function symbols},
\emph{predicate symbols}, \emph{variables} and \emph{labels} respectively.
A \emph{symbol} means an  element of $ {\f} \cup {\p}$ in this paper.
For example, the schema in Figure \ref{S.fig} has function set $\f = \{f, g, h\}$,
predicate set $\p = \{p\}$ and variable set $\var = \{u, v\}$.
We assume  a  function
$$\arity: \f \cup\p \to \Bbb{N}.$$
The arity of a symbol $x$ is the number of arguments referenced
by $x$, for example in the schema in Figure \ref{S.fig} the function $f$ has arity one,
the function $g$ has arity zero, and $p$ has arity one.

 Note that in the case when the arity of a function symbol $g$ is zero,
 $g$ may be thought of as a constant.

 The set $ \term({\f}, {\var})$ of \emph{terms} is defined as follows:
\begin{itemize}
\item each variable is a term,
\item
if $ f\in {\f}$ is of arity $
n$ and $ t_1,\ldots ,t_n$ are terms then $ f(t_1,\ldots ,t_n)$ is a term.
\end{itemize}

For example, in the schema in Figure \ref{S.fig}, the variable $u$ takes the value (term) $h();$ after the first assignment is executed and if we take the true branch then the variable $v$ ends with the value (term) $f(h())$.

We refer to a tuple $\vt =(t_1,\ldots, t_n)$, where each $t_i$ is a term, as a vector term. We call $p(\vt)$ a predicate term if $p\in \p$
and the number of components of the  vector term $\vt$  is $\arity(p)$.

Schemas are defined recursively as follows.

\begin{itemize}
\item
$\ski$ is a schema.
\item Any label is a schema.
\item
An \ass\ $y\as f(\vx);$ for a variable $y$, a function symbol $f$ and an $n$-tuple $\vx$ of variables, where $n$ is the arity of $f$, is a schema.
\item
If $S_1$ and  $S_2$ are  schemas then $S_1S_2$ is a schema.
\item
If $S_1$ and  $S_2$ are  schemas, $p$ is a predicate symbol and $\vy$ is an $m$-tuple of variables, where $m$ is the arity of $p$, then $\si p(\vy) \then S_1 \els S_2$ is a schema.
\item
If $T$ is a schema, $q$ is a predicate symbol and $\vz$ is an $m$-tuple of variables, where $m$ is the arity of $q$,  then  the schema $\whi q(\vz)\;T$ is a schema.
\end{itemize}

If no function or predicate symbol, or label, occurs more than once in a schema $S$, we say that $S$ is linear. If a schema does not contain any predicate symbols, then we say it is predicate-free. If a linear schema $S$ contains a subschema $\si p(\vy) \then S_1 \els S_2$, then
we refer to $S_1$ and $S_2$ as the $\tru$-part and $\fal$-part respectively of $p$ in $S$. For example in the schema in Figure \ref{S.fig} the predicate $p$ has $\tru$-part $v\as f(u);$ and $\fal$-part $v\as g();$.
If a linear schema $S$ contains a subschema $\whi q(\vz)\;T$, then  we refer to $T$ as the body of $q$ in $S$.

Quotients of schemas are defined recursively as follows; $\ski$ is a quotient of every schema; if $S'$ is a quotient of $S$ then $S'T$ is a quotient of $ST$ and $TS'$ is a quotient of $TS$;  if $T'$ is a quotient of $T$, then $\whi q(\vy)\;T'$ is a quotient of $\whi q(\vy)\;T$; and if $T_1$ and $T_2$ are quotients of schemas  $S_1$ and $S_2$ respectively, then $\si p(\vx) \then T_1 \els T_2$ is a quotient of $\si p(\vx) \then S_1 \els S_2$. A quotient $T$ of a schema $S$ is said to be non-trivial if $T\not= S$.

Consider the schema in Figure \ref{S.fig}. Here we can obtain a quotient by replacing the first statement by $\ski$ or by replacing the if statement by $\ski$. It is also possible to replace either or both parts of the if statement by $\ski$ or any combination of these steps.

\subsection{Paths through a schema} \label{path.defn.subsection}

We will express the semantics of schemas using paths through them; therefore  the definition of a path through a schema has to include the  variables assigned or referenced by successive function or predicate symbols.

The set of prefixes of a \emph{word} (that is, a sequence) $\sigma$ over an alphabet is denoted by $\pre(\sigma)$.
For example, if $\sigma =x_1 x_3 x_2$ over the alphabet $\{x_1,x_2,x_3\}$, then the set $\pre(\sigma)$ consists of the words $x_1x_2x_3,x_1x_2,x_1$ and the empty word. More generally, if $\Omega$ is a set of words, then we define $\pre(\Omega)= \{\pre(\sigma)\vert\; \sigma \in \Omega\}$.

For each schema $S$ there is an associated  alphabet $\lang(S)$ consisting of all elements of $\mathcal L$ and   the set of letters of the form  $ \ul{y\as f(\vx)}$ for \asss\
$y\as f(\vx);$ in $S$  and $\ul{p(\vy),Z}$ for  $Z\in \tf$, where $\si p(\vy)$ or $\whi p(\vy)$ occurs in $S$. For example, the schema in Figure \ref{S.fig} has no labels and has alphabet \\
$\{ \ul{y\as h()},  \ul{v\as f(\vu)},  \ul{v\as g()}, \ul{p(\vw),\tru},  \ul{p(\vw),\fal}\}$.
The set $\pathset(S)$ of terminating paths through $S$, is defined recursively as follows.

\begin{itemize}
\item
$\pathset(l)=l$, for any  $l \in \mathcal L$.
\item
$\pathset(\ski)$ is the empty word.
\item
$\pathset(y\as f(\vx);)=\ul{y\as f(\vx)}$.
\item
$\pathset(S_1S_2)=\pathset(S_1)\,\pathset(S_2)$.
\item
$\pathset(\si p(\vx) \then S_1 \els S_2)=\, \ul{p(\vx),\tru}\, \pathset(S_1) \cup\, \ul{p(\vx),\fal}\, \pathset(S_2)$.
\item
$\pathset(\whi(q(\vy) \, T))=\,  (\ul{q(\vy),\tru}\,\pathset(T))^*\,\ul{q(\vy),\fal}$.
\end{itemize}

We sometimes abbreviate $\ul{q(\vy),Z}$ to $\ul{q,Z}$ and $\ul{y\as f(\vx)}$ to $\ul{f}$.

We define $\allpath(S)$ to be the set containing $\pathset(S)$, plus all infinite words 
 whose finite prefixes are  prefixes  of  terminating paths.
A \emph{path} through $S$ is any (not necessarily strict) prefix of an element of    $\allpath(S)$. As an example, if 
$S$ is the schema in 
 Figure \ref{S.fig}, which  has no loops, then $\pathset(S)=\allpath(S)$.
 In fact, $\pathset(S)$ in this case contains exactly two paths, defined by $p(\vw)$ taking the true or false branches, and every path through $S$ is  a prefix of one of these paths.

If $S'$ is a quotient of a schema $S$, and $\rho \in \pre(\pathset(S))$ (that is, $\rho$ is a path through $S$), then $\proj[S'](\rho)$ is the path obtained from $\rho$ by deleting all letters having function or predicate symbols  not lying in $S'$ and all labels not occurring in $S'$. It is easily proved that $\proj[S'](\pathset(S))= \pathset(S')$ in this case.

 \subsection{Semantics of   schemas} \label{struc.sch.sem.subsec}

The symbols upon which schemas are built are given meaning by
defining the notions of a state and of an interpretation. It
will be assumed that `values' are given in a single set $D$,
which will be called the \emph{domain}. We are mainly interested in the case in which $D=\term(\f,\var)$ (the Herbrand domain) and the function symbols represent the `natural' functions \wrt\ $\term(\f,\var)$.

 \begin{defn}[states, (Herbrand) \ters\ and the natural state $e$] \rm
 \label{state.ter.defn}
\mbox{}\\
Given a domain $D$, a \emph{state} is either $\bot$ (denoting non-termination) or a function $\var \ra D$. The set
of all such states
will be denoted by $\state(\var,D)$.
An interpretation $i$ defines, for each function symbol
$f\in\f$ of arity $n$, a function $ f^i:D^n \ra D$, and for each
predicate symbol $p\in\p$ of arity $m$, a function
$ p^i:D^m \ra \{\tru ,\, \fal \}$. The set of all
interpretations with domain $D$ will be denoted
$\Int({\f}, {\p}, D)$.

We call the set $\term(\f,\var)$ of terms the \emph{Herbrand domain}, and we say that a function from $\var$ to $\term(\f,\var)$ is a Herbrand state.
An interpretation $i$ for the Herbrand domain is said to be \emph{Herbrand}   if the functions $f^i:\, \term(\f,\var)^n \to \term(\f,\var)$ for each $f\in \f$
are defined as
\begin{center}$f^i(t_1,\ldots,t_n)= f(t_1,\ldots,t_n)$ \end{center}
for all $n$-tuples of terms $(t_1,\ldots,t_n)$.

We define
the \emph{natural state} $e:\var \ra \term(\f,\var)$
 by $e(v)=v$ for all $ v\in \var.$
\end{defn}

In the schema in Figure \ref{S.fig} the natural state simply maps variable $\vu$ to the name $u$, variable $\vv$ to the name $v$, and variable $\vw$ to the name $w$. The program in Figure \ref{program.fig} can be produced from this schema through the interpretation that maps $h();$ to $1$, $p(\vw)$ to $\vw>1$, $f(\vu)$ to $\vu+1$, and $g()$ to $2$; clearly this is not a Herbrand interpretation.

Observe  that if  an \ter\ $i$ is   Herbrand, this does not restrict   the mappings \\
$ p^i:(\term({\f}, \var))^m \ra \{\tru ,\, \fal \}$ defined by   $ i$ for each $p\in \p$.

It is well known ~\cite[Section 4-14]{manna:book}
that Herbrand \ters\ are
the only ones that need to be considered when considering many
schema properties.
This fact is stated more precisely in Theorem~\ref{freeint.thm}. In particular, our semantic slicing definitions may be
 defined in terms of Herbrand domains.

Given a schema $S$ and a domain $D$, an
initial state $d\in\state(\var,D)$ with $d\not= \bot$ and an interpretation
$i\in\Int({\f}, {\p}, D)$ we now define the final state
$\dd{S}{i}\in\state(\var,D)$ and the associated   path
$\path{S}{i,d}\in\allpath(S)$. In order to do this, we need to define the predicate-free schema associated with the prefix of a path by considering the sequence of \asss\ through which it passes.


\begin{defn}[the schema $\schema(\sigma)$]\rm \mbox{}\\ \label{schema.of.path.defn}
Given a word $ \sigma \in (\lang(S))^*$ for a schema $S$, we recursively define  the predicate-free
schema $\schema(\sigma)$ by the following rules;  $\,\schema(\ski)=\ski$, $\,\schema(l)=l$ for  $l \in \mathcal L$,
 $\schema(\sigma \ul{ v\as {f}(\vx)}) \;= \; \schema(\sigma)\, v\as {f}(\vx);$ and $\schema(\sigma \ul{p(\vx),X})\; = \; \schema(\sigma)$.
\end{defn}

Consider, for example, the path of the schema in Figure \ref{S.fig} that passes through the true branch of $p$. Then this defines a word $\sigma = \ul{ u \as {h}()} \ul{p(\vw),\tru} \ul{ v \as {f}(\vu)} $ and $\schema(\sigma) = \ul{ u \as {h}()} \ul{ v \as {f}(\vu)}$.

\begin{lem}
\label{exe.lem}
Let $S$ be a schema.
 If  $\sigma \in \pre(\pathset(S))$,  the set
$\{ m\in \lang(S) \vert\, \sigma m \in \pre(\pathset(S)) \}$ is
one of the following; a label,  a singleton containing  an assignment letter $\ul{y\as f(\vx)}$,  a
 pair  $\{ \ul{ {p}(\vx) ,\tru} ,\;\ul{ {p}(\vx),\fal} \}$ for a predicate $p$ of $S$, or the empty set, and if $\sigma \in \pathset(S)$ then the last case holds.
\end{lem}

\proof  \cite[Lemma 6]{laurence:flfl}. \proofend

Lemma \ref{exe.lem}
reflects the fact that at any point in the execution of a program,
there is never more than one `next step' which may be taken, and an element of $\pathset(S)$ cannot be a strict prefix of another.


 \begin{defn} [semantics of predicate-free schemas]
 \rm \label{meaning}
Given a state $d\not= \bot$,  the final state $\dd{S}{i}$ and associated path $\path{S}{i,d} \in \allpath(S)$
of a schema $S$ are defined
as follows:\\
\begin{itemize}
\item $\dd{\ski}{i}= d$ and $\path{\ski}{i,d}$ is the empty word.
\item $\dd{l}{i}= d$ and $\path{l}{i,d}=l$  for $l \in \mathcal L$.
\item $ \dd{y \as f(\vx);}{i} (v)~~~=~~~
\begin{cases} d(v) & \text{if $ v \not= y$},
\\  f^{i}(d(\vx))  & \text{if $ v=y$ }
 \end{cases}$
 (where the vector term $d(\vx)=(d(x_1),\ldots, d(x_n))$ for $\vx=(x_1,\ldots, x_n)$), and
\\ \mbox{}\\
$\path{y\as f(\vx);}{i,d}~~~=~~~  \ul{ y\as f(\vx)}$. \\
\item
For sequences $S_1S_2$  of predicate-free schemas,
$\dd{S_1S_2}{i}~~~=~~~{\mathcal M}\quine{S_2}^{i}_{\dd{S_1}{i}}$ and
\\ \mbox{}\\
$\path{S_1S_2}{i,d}~~~=~~~\path{S_1}{i,d}\path{S_2}{i,\dd{S_1}{i}}$.
\end{itemize}
\end{defn}

This uniquely defines $\dd{S}{i}$ and $\path{S}{i,d}$ if $S$ is predicate-free.
 In order to give the semantics of a general schema  $S$,
 first the path, $ \path{S}{i,d}$, of $S$
 with respect to interpretation, $i$, and
 initial state $d$ is defined.

\begin{defn}[the path $ \path{S}{i,d}$]
\rm
Given  a schema $S$, an interpretation $i$, and a
state, $d\not= \bot$, the path
 $  \path{S}{i,d} \in  \allpath(S)$ is defined by the   following condition;  for all
\\
 $\sigma \, \ul{ p(\vx), Z} \; \in \pre(\path{S}{i,d})$, the equality  $p^i(\dd{\schema(\sigma)}{i} (\vx))=Z$ holds.
 \end{defn}

In other words, the path   $ \path{S}{i,d}$
has the following property;  if a predicate expression $p(\vx)$ along  $ \path{S}{i,d}$  is evaluated
with respect to the predicate-free schema consisting of the
sequence of assignments preceding  that predicate in $ \path{S}{i,d}$, then the
value of  the resulting predicate term given by $i$ `agrees' with
the value given in $ \path{S}{i,d}$. Consider, for example, the schema given in Figure \ref{S.fig}
and the interpretation that gives the program in Figure \ref{program.fig}. Given a state $d$ in which $w$
has a value greater than one, we obtain the path $\ul{ u \as {h}()}\; \ul{p(\vw),\tru}\; \ul{ v \as {f}(\vu)} $.

By   Lemma~\ref{exe.lem}, this defines the path
$\path{S}{i,d} \in \allpath(S)$ uniquely.

\begin{defn}[the semantics of arbitrary schemas] \label{genmeaning}
\rm
 If
$\path{S}{i,d}$ is finite, we define
$$\dd{S}{i}=\dd{\schema(\path{S}{i,d})}{i}$$
(which is already defined, since
$ \schema(\path{S}{i,d})$ is predicate-free) otherwise
$\path{S}{i,d}$ is infinite  and we define $\dd{S}{i}=\bot$. In this last case we may say that $\dd{S}{i}$ is not terminating.

For convenience, if $S$ is predicate-free and $ d: \var \ra \term(\f,\var)$ is a state
then we define unambiguously
$ \dd{S}{}=\dd{S}{i}$; that is, we assume that the \ter\ $i$ is Herbrand if  $d$ is a Herbrand state. Also, if $\rho$ is a path through a schema, we may write $\ee{\rho}{}$ to mean $\ee{\schema(\rho)}{}$.
\end{defn}

Observe that $\dd{S_1S_2}{i}={\mathcal M}\quine{S_2}^{i}_{\dd{S_1}{i}}$ and
$$\path{S_1S_2}{i,d}=\path{S_1}{i,d}\path{S_2}{i,\dd{S_1}{i}}$$ hold for all schemas (not just predicate-free ones).

Given a  schema $S$ and  $\mu \in \pre(\pathset(S))$, we say that $\mu$ passes through a predicate term $p(\vt)$  if $\mu$ has a prefix $\mu'$ ending  in $\ul{ {p}(\vx), Y}$ for $y\in \tf$  \st\ $\mean{\schema(\mu')}{}{e}(\vx) = \vt$ holds. In this case  we say that $p(\vt)= Y$ is  a \emph{consequence} of $\mu$.
For example, the path $\ul{ u \as {h}()}\; \ul{p(\vw),\tru}\; \ul{ v \as {f}(\vu)} $ of the schema in Figure \ref{S.fig} passes through the predicate term $p(\vw)$ since this path has no assignments to $w$ before $p$.


\begin{defn}[path compatibility and executability] \rm \label{path.compat.execute.defn}
Let $\rho$ be a path through a schema $S$. Then $\rho$ is \emph{executable} if $\rho$ is a prefix of $\path{S}{j,d}$ for some \ter\ $j$ and state $d$. Two paths $\rho,\rho'$ through schemas $S,S'$ are \emph{compatible}  if for some \ter\ $j$ and state $d$, they are prefixes of $\path{S}{j,d}$ and $\path{S'}{j,d}$ respectively.
\end{defn}

The justification for restricting ourselves to consideration of Herbrand \ters\ and the state $e$ as the initial state lies in the fact that Herbrand \ters\ are the `most general' of \ters.\ Theorem  \ref{freeint.thm}, which is virtually a restatement  of \cite[Theorem 4-1]{manna:book},  expresses this  formally.

\begin{thm}
\label{freeint.thm}
Let $\chi $ be a set of schemas, let $ D$ be a domain, let $d$ be a function from the set of variables into $D$ and let $i$ be an \ter\ using this domain.  Then there is a Herbrand \ter\ $j$ \st\ the following hold.
\begin{enumerate}
\item
 For all $S\in \chi$, the path $\path{S}{j,e}=\path{S}{i,d}$.
\item
If $S_1,S_2 \in \chi$ and $v_1,v_2$ are variables and  $\rho_k \in \pre(\path{S_k}{j,e})$ for $k=1,2$ and $\ee{\rho_1}{} (v_1)=\ee{\rho_2}{} (v_2)$, then also  $\dd{\rho_1}{i} (v_1)=\dd{\rho_2}{i} (v_2)$ holds.
\end{enumerate}
\end{thm}

As a consequence of Part (1) of Theorem \ref{freeint.thm}, it  may be assumed in Definition \ref{path.compat.execute.defn} that $d=e$ and the \ter\  $j$ is   Herbrand without strengthening the Definition. In the remainder of the paper we will assume that all \ters\ are Herbrand.


\section{The path-faithful dynamic slicing criterion}   \label{dyn.defn.sect}

In this section we adapt the notion of a dynamic program slice to program schemas.
Dynamic program slicing is formalised in the original paper by
Korel and Laski \cite{korel:dynamic-slicing}. Their definition uses two
functions, $Front$ and $DEL$, in which $Front(T,i)$ denotes the
first $i$ elements of a trajectory\footnote{A trajectory is a path in which we do
not distinguish between true and false values for a predicate. There is a one-to-one
correspondence between paths and trajectories unless there is an if statement that contains
only $\ski$.} $T$ and $DEL(T,\pi)$ denotes
the trajectory $T$ with all elements that satisfy predicate $\pi$
removed. A trajectory is a path through a program, where each node
is represented by a line number and so for path $\rho$ we have
that $\hat{\rho}$ is the corresponding trajectory.

Korel and Laski use a slicing criterion that is a tuple $c = (x,I^q,V)$ in which $x$
is the program input being considered, $I^q$ denotes the execution
of statement $I$ as the $q$th statement in the path taken when $p$
is executed with input $x$, and $V$ is the set of variables of interest.

The following is the definition provided\footnote{Note that this
almost exactly a quote from \cite{korel:dynamic-slicing} and is taken from
\cite{BinkleyDGHKK06}}:

\begin{defn}
Let $c = (x, I^q, V)$ be a slicing criterion of a program $p$
and $T$ the trajectory of $p$ on input $x$. A dynamic slice of $p$
on $c$ is any executable program $p'$ that is obtained from $p$ by
deleting zero or more statements such that when executed on input
$x$, produces a trajectory $T'$ for which there exists an
execution position $q'$ such that

\begin{itemize}
\item (KL1) $Front(T', q') = DEL(Front(T , q), T (i) \not \in N'
\wedge 1 \leq i \leq q)$,

\item (KL2) for all $v \in V$, the value of $v$ before the
execution of instruction $T (q)$ in $T$ equals the value of $v$
before the execution of instruction $T'(q')$ in $T'$,

\item (KL3) $T'(q') = T(q) = I$,
\end{itemize}
where $N'$ is a set of instructions in $p'$.
\end{defn}

In producing a dynamic slice all we are allowed to do is to
eliminate statements. We have the requirement that the slice and
the original program produce the same value for each variable in
the chosen set $V$ at the specified execution position and
that the path in $p'$ up to $q'$ followed by using input $x$ is
equivalent to that formed by removing from the path $T$ all
elements not in the slice. Interestingly, it has been observed
that this additional constraint, that $Front(T', q') = DEL(Front(T
, q), T (i) \not \in N' \wedge 1 \leq i \leq q)$, means that a
static slice is not necessarily a valid dynamic slice
\cite{BinkleyDGHKK06}.

We can now give a corresponding definition for linear schemas.

\begin{defn}[path-faithful dynamic slice]\rm \label{dyn.path.slice.defn}
Let $S$ be a linear schema containing a label $l$, let $V$ be a set of variables and let  $\rho\, l\in \pre(\pathset(S))$ be executable.  Let $S'$ be a quotient of $S$ containing $l$. Then we say that $S'$ is a $(\rho l,V)$-path-faithful dynamic slice (PFDS) of $S$ if
the following hold.
\begin{enumerate}
\item
Every variable in $V$ defines the same term after $\proj(\rho)$  as after $\rho$ in $S$.
\item
Every maximal path through $S'$ which is compatible with $\rho$ has  $\proj(\rho)$ as a prefix.
\end{enumerate}
\end{defn}

If the label $l$ occurs at the end of $S$, so that $S= T \,l $ for a schema $T$, and $S'$ is a $(\rho l,V)$-dynamic  slice of $S$, so that $S'=T' \, l$, then we simply say that $T'$ is a $(\rho ,V)$-path-faithful dynamic  end slice of $T$.

\begin{thm} \label{dps.verify.thm}
Let $S$ be a   linear schema, let   $\rho l\in \pre(\pathset(S))$ be executable, let $V$ be a set of variables and let $S'$ be a quotient of $S$ containing $l$. Then $S'$ is a $(\rho l,V)$-PFDS of $S$ \ifa\ $\ee{\rho}{}(v)=  \ee{\proj(\rho)}{}(v)$ for all $v\in V$ and every expression $p(\vt)=X$ which is a consequence of $\proj(\rho)$ is also  a consequence of $\rho$.
\end{thm}

\proof
This follows immediately from the two conditions in Definition
\ref{dyn.path.slice.defn}.
\proofend

As an example of a path-faithful dynamic  end slice, consider the linear schema of Figure
\ref{distinct.fig}. We assume that $V= \{v\}$ and the path
$$\rho=\, (\,\ul{p,\tru}\; \ul{g}\;\ul{f}\;\ul{q,\tru}\; \ul{h}\; \ul{H}\,)^2\,\ul{p,\fal}$$
 which passes twice through the body of $p$, in each case passing through $\ul{q,\tru}$, and then leaves the body of $p$. Thus the  value of $v$ after $\rho$ is $f(h(u))$. Thus any $(\{v\},\rho)$-DPS $S'$ of $S$ must contain $f$ and $h$ in order that ($1$) is satisfied,  and hence  contains $p$ and $q$. By Theorem \ref{dps.verify.thm}, $S'$ would also have to contain $g$, since otherwise $p(w)=\fal$ would be a consequence of $\proj(\rho)$, whereas $p(w)=\fal$ is not a consequence of $\rho$. Also, $S'$ would contain  the function symbol $H$, since otherwise $q(g(w),t)=\tru $ would be a consequence of $\proj(\rho)$, but not of $\rho$. Thus $S$ itself is the only $(\{v\},\rho)$-PFDS  of $S$. Observe that the inclusion of the \ass\ $t\as H(t);$ has the sole effect of ensuring that for every \ter\ $i$ for which $\path{S'}{i,e}=\rho$,
 $\path{S'}{i,e}$ passes through $\ul{q,\tru}$ instead of $\ul{q,\fal}$ during its second passing through the body of $p$, and so deleting $t\as H(t);$ does not alter the value of $v$ after $\path{S'}{i,e}$. This suggests that our definition of a dynamic  slice may be unnecessarily restrictive, and this motivates the generalisation of Definition \ref{gen.dyn.path.slice.defn}.

\begin{figure}[h]

\begin{center}
$
\begin{array}{llll}
\whi p(w)& \{\\
& w\as g(w);  \\
    & v \as f(u);  \\
& \si q(w,t) \then & u\as h(u);\\
& t\as H(t);\\
&\}\\
 \end{array}
 $
\end{center}
\caption{A linear schema with distinct minimal dynamic and path-faithful dynamic  slices} \label{distinct.fig}
\end{figure}

\section{A New Form of Dynamic Slicing} \label{gen.dyn.defn.sect}

Path-faithful dynamic slices of schemas correspond to dynamic program slices and in order to
produce a dynamic slice of a program we can produce the path-faithful dynamic slice of the
corresponding linear schema. In this section we show how this notion of dynamic slicing can be
weakened, to produce smaller slices, for linear schemas and so also for programs.

Consider the schema in  Figure \ref{distinct.fig}, the path $\rho = \,\ul{p,\tru}\, \ul{g}\,\ul{f}\,\ul{q,\tru}\,\ul{h}\,\ul{H}\,\ul{p,\tru}\, \ul{g}\,\ul{f}\,\ul{q,\tru}\,\ul{h}\,\ul{H}\,\ul{p,\fal}$ and variable $v$. It is straightforward to see that a dynamic slice has to retain the predicate $p$ since it controls a statement ($u\as h(u)$) that updates the value of $u$ and this can lead to a change in the value of $v$ on the next iteration of the loop. Thus, a dynamic slice with regards to $v$ and $\rho$ must retain predicate $q$. Further, the assignment $t \as H(t)$ affects the value of $t$ and so the value of $q$ on the second iteration of the loop in $\rho$ and so a (path-faithful) dynamic slice must retain this assignment.

We can observe that in $\rho$ the value of the predicate $q$ on the last iteration of the loop does not affect the final value of $v$. In addition,
in $\rho$ the assignment $t \as H(t)$ only affects the value of $q$ on the last iteration of the loop and this assignment does not influence the final value of $v$. In this section we define a type of dynamic slice that allows us to eliminate this assignment. At the end of this section we describe a context in which we might be happy to eliminate such assignments.

\begin{prop}\label{substitute.prop}
Let $S$ be a linear schema and let $\rho$ be a path through $S$.
\begin{enumerate}
\item
Let $q$ be a while predicate in $S$ and let $\mu$ be a terminal path in the body of $q$ in $S$. Then a word $\alpha \ul{q,\tru} \mu\ul{q,\fal}\gamma$ is a path in $S$ \ifa\ $\alpha \ul{q,\fal}\gamma$ is a path in $S$.
\item
Let $q$ be an if  predicate in $S$, let $Z\in \tf$ and let $\mu,\mu'$ be  terminal paths in the $Z$-part and $\neg Z$-part respectively  of $q$ in $S$. Then a word $\alpha \ul{q,Z} \mu\gamma$ is a path in $S$ \ifa\ $\alpha \ul{q,\neg Z}\mu'\gamma$ is a path in $S$.
\end{enumerate}
Furthermore, in both cases, one path is terminal \ifa\ the other is terminal.
\end{prop}

\proof
Both  assertions follow straightforwardly by structural induction from the definition of $\pathset(S)$ in Section  \ref{path.defn.subsection}.
\proofend

\begin{defn}\label{reducible.defn}\rm
Let $S$ be a linear schema, let $l$ be a label  and let $\rho ,\rho'$ be paths through $S$. Then we say that $\rho$ is simply $l$-reducible to $\rho'$ if $\rho'$ can be  obtained from  $\rho $ by one of the following transformations, which we call simple $l$-reductions.
\begin{enumerate}
\item
Replacing a segment $\ul{p,\tru} \,\sigma \, \ul{p,\fal}$ within $\rho$ by  $\ul{p,\fal}$, where $\sigma$ is a terminal path in the body of a while predicate $p$ which does  not contain $l$ in its body.
\item
Replacing a segment $\ul{p,Z} \,\sigma$ within $\rho$ by  $\ul{p,\neg Z}$, where $\sigma$ is a terminal path in the $Z$-part of an if predicate $p$,  $l$ does not lie in either part of $p$ and the $\neg Z$-part of $p$  is $\ski$.
\end{enumerate}
If $\rho'$ can be  obtained from  $\rho $ by applying zero or more $l$-reductions, then we say that $\rho$ is  $l$-reducible to $\rho'$. If the condition on the  label $l$ is removed from the definition  then we use the terms reduction and simple reduction.
\end{defn}

By Proposition \ref{substitute.prop}, the transformations given in Definition \ref{reducible.defn} always produce paths through $S$.
Observe that if  $\rho$ is  $l$-reducible to $\rho'$, then the sequence of function and predicate symbols through which $\rho'$ passes is a subsequence of that through which $\rho$ passes,
$\rho$ and  $\rho'$ pass through the label $l$ the same number of times, and the length of $\rho'$ is not greater than that of $\rho$.

\begin{defn}[dynamic  slice]\rm \label{gen.dyn.path.slice.defn}
Let $S$ be a linear schema containing a label $l$, let $V$ be a set of variables and let $\rho\, l\in \pre(\pathset(S))$ be executable.  Let $S'$ be a quotient of $S$ containing $l$. Then we say that $S'$ is a $(\rho l,V)$-dynamic  slice (DS) of $S$ if every maximal path through $S'$ compatible with $\rho$  has a prefix  $\rho'$ to which $\proj(\rho) $ is $l$-reducible    and \st\  every variable in $V$ defines the same term after $\rho'$  as after $\rho$ in $S$.
\end{defn}

If the label $l$ occurs at the end of $S$, so that $S= T \,l $ for a schema $T$, and $S'$ is a $(\rho l,V)$-dynamic  slice of $S$, so that $S'=T' \, l$, then we simply say that $T'$ is a $(\rho ,V)$-dynamic  end slice of $T$.

Consider again the schema in Figure
\ref{distinct.fig} and path
$\rho = \, \ul{p,\tru}\, \ul{g}\, \ul{f}\,\ul{q,\tru}\, \ul{h}\, \ul{H}\, \ul{p,\tru}\, \ul{g}\, \ul{f}\, \ul{q,\tru}\, \ul{h}\, \ul{H}\, \ul{p,\fal}$.
Here the quotient $T$ obtained from $S$ by deleting the \ass\ $t\as H(t);$ is a $(\rho,v)$-dynamic  end slice of $S$, since the path
$$\rho'=\, \ul{p,\tru}\, \ul{g}\,\ul{f}\,\ul{q,\tru}\, \ul{h}\, \ul{H}\, \ul{p,\tru}\, \ul{g}\, \ul{f}\, \ul{q,\fal}\, \ul{p,\fal}$$
 is simply reducible from $\proj(\rho)$ and gives the correct final value for $v$, and $\rho'$ and $\proj(\rho)$ are the only maximal paths through $S'$ that are  compatible with $\rho$.
This shows that  a DS of a linear schema may be  smaller than a PFDS.

One area in which it is useful to determine the dependence along a
path in a program is in the application of test techniques,
such as those based on evolutionary algorithms, that
automate the generation of test cases to satisfy a structural criterion.
These techniques may choose a path to the point of the program to be covered
and then attempt to generate test data that follows the path (see, for example, \cite{HarmanFHHDW02,jones96,wegener96,wegener97}).
If we can determine the inputs that are relevant to this path then we can focus on these
variables in the search, effectively reducing the size of the search space.
Current techniques use static slicing but there is potential for using dynamic slicing
in order to make the dependence information more precise and, in particular,
the type of dynamic slice defined here.

\begin{figure}[h]

\begin{center}
$
\begin{array}{llllll}
\whi P(v)& \{\\
&\si  Q(v) & \then & \{ \\
 &&& \si q(v) \then & \{ \\
                       &&&&  x \as g_{good}();\\
                        &&&& v \as G_{good}(x,v);\\
                      &&&& \} \\
&&& \;\;\;\;\;\;\;\;\;\;\;\;\els & \{ \\
                       &&&&  x \as g_{bad}();\\
                       &&&& v \as G_{bad}(x,v);\\
                      &&&& \} \\
\\

&&& \si s_1(v) \then & x \as g_1();\\
&&& \si s_2(v) \then &  x \as g_2();\\
    \\
&&& \si t(x) \;\;\;\; \then & v\as H(v); \\
\\
&&& \}\\
&& \els       &   \ski; \\
& v \as J(v);\\
&\}\\
 \end{array}
 $
\end{center}
\caption{A linear  schema with distinct minimal path-faithful dynamic slices} \label{two.mins.gdps.fig}
\end{figure}

\section{A  linear schema with two minimal path-faithful dynamic slices} \label{two.mins.sect}

Given a linear schema, a variable set $V$ and a path $\rho$ through $S$, we wish to establish information about the set of all $(\rho,V)$-dynamic  slices, which is partially ordered by set-theoretic inclusion of function and predicate symbols. In particular, it would be of interest to obtain conditions on $S$ which would ensure that minimal slices were unique since under such conditions it may be feasible to produce minimal slices in an incremental manner, deleting one statement at a time until no more statements can be removed. As we now show, however, this is false for  arbitrary linear schemas, whether or not slices are required to be path-faithful. To see this,
consider the schema $S$ of Figure \ref{two.mins.gdps.fig} and the slicing criterion defined by the variable $v$ and  the terminal  path $\rho$  which  enters the body of $P$ 5 times as follows.

1st time; $\rho$ passes through $g_{good}$ and $H$, but not through either $g_i$.\\
2nd time; $\rho$ passes through $g_{good}$, $g_1$ and $H$, but not through  $g_2$.\\
3rd time; $\rho$ passes through $g_{good}$, $g_2$ and $H$, but not through  $g_1$.\\
4th time; $\rho$ passes through $g_{bad}$, $g_1$, $g_2$ and $H$. \\
5th time; $\rho$ passes through $\ul{Q,\fal}$.

Define the quotient $S_1$  of $S$  by deleting the entire if statement guarded by  $s_2$ and define $S_2$  analogously by interchanging the suffices 1 and 2.  By Theorem \ref{dps.verify.thm},  $S_1$ and $S_2$ are both $(\rho,v)$-PFDS's of $S$, since $t(x)$ will still evaluate to $\tru$ over the path $\proj[S_1](\rho)$ or $\proj[S_2](\rho)$ on paths 2--4.  \Otoh,\ if the  if statements guarded by $s_1$ and   $s_2$ are both  deleted, then on the 4th path, $t(x)$ may evaluate to $\fal$, since $g_{bad}$ never occurs in the predicate term defined by $t(x)$ along $\rho$, hence the final value of $v$ may contain fewer occurrences of $H$ in the slice than after $\rho$.  Furthermore, every $(\rho,v)$-DS of $S$ must contain the function symbols $J$, $H$, $G_{good}$ and $G_{bad}$ and hence $g_{good}$ and $g_{bad}$, since the final term defined by $v$ contains these symbols, and so $S_1$ and $S_2$ are minimal $(\rho,v)$-DS's, and are also both path-faithful.

\section{Decision problems for dynamic  slices} \label{decide.dps.sect}

In this section, we establish complexity bounds for two problems; whether a quotient $S'$ of a linear schema $S$ is a dynamic  slice, and whether a linear schema $S$ has a non-trivial dynamic  slice. We consider the problems both with and without the requirement that dynamic slices be path-faithful.


\begin{defn}[maximal common prefix of a pair of words] \rm
 The maximal common prefix of words  $\sigma,\sigma'$ is denoted by $\maxpre(\sigma,\sigma')$. For example, the maximal common prefix of the words $x_1x_2x_3x_4$ and $x_1x_2yx_4$ over the five-word alphabet $\{x_1,x_2,x_3,x_4,y\}$ is $x_1x_2$; that is, $\maxpre(x_1x_2x_3x_4,x_1x_2yx_4)=x_1x_2$.
\end{defn}

\begin{lem}\label{reducible.path.order.lem}
Let $S$ be a linear schema containing a label $l$ and let $\rho ,\rho'$ be paths through $S$. Suppose  $\rho$ is $l$-reducible to $\rho'$.  Then there is a sequence $\rho_1=\rho,\ldots, \rho_n=\rho'$ \st\ each $\rho_{i}$ is simply $l$-reducible to $\rho_{i+1}$, and $\maxpre(\rho_{i},\rho_{i+1})$ is always a strict prefix of $\maxpre(\rho_{i+1},\rho_{i+2})$. 
\end{lem}

\proof
This follows from the fact that the two transformation types commute. Since  $\rho$ is $l$-reducible to $\rho'$,  there is a sequence $\rho_1=\rho,\ldots, \rho_n=\rho'$  \st\ each $\rho_i$ is obtained from $\rho_{i-1}$ by a simple $l$-reduction, and we may assume that $n$ is minimal. Thus for each $i <n$, and using the definition of a simple $l$-reduction, we can  write
$\rho_i= \alpha_i \ul{p_i,Z_i}\beta_i \gamma_i$ and  $\rho_{i+1}= \alpha_i \ul{p_i,\neg Z_i} \gamma_i$. If every $\alpha_i$ is a strict prefix of $\alpha_{i+1}$, then the sequence of  paths $\rho_i$ already satisfies  the required property. Thus \wma\ that for some minimal $i$, $\alpha_{i}$ is not
 a strict prefix of $\alpha_{i+1}$.

We now compare the two ways of writing $$\rho_{i+1}= \alpha_i \ul{p_i,\neg Z_i} \gamma_i=\alpha_{i+1} \ul{p_{i+1}, Z_{i+1}}\beta_{i+1} \gamma_{i+1}.$$ Clearly $\alpha_{i+1}$ is a prefix of $\alpha_{i}$. We consider three cases.
\begin{enumerate}
\item
Suppose that $\alpha_{i}=\alpha_{i+1}$.   Thus the first letter of $\rho_{i+1}$ after $\alpha_i$ is $\ul{p_i,\neg Z_i}= \ul{p_{i+1}, Z_{i+1}}$. If $p_i=p_{i+1}$ were a while predicate, then $Z_i=\tru$ would follow from the fact that $\rho_{i}$ is $l$-reducible to $\rho_{i+1}$,
 and $Z_{i+1}=\tru$ would follow similarly from the pair $\rho_{i+1},\rho_{i+2}$, giving a contradiction, hence
 $\,p_i$ must be an if predicate and so   the $\neg Z_i$-part and the $Z_i=\neg Z_{i+1}$-part of $p$ is  $\ski$  from the definition of $l$-reduction
 and hence $\rho_{i+2}=\rho_i$ holds, contradicting the minimality of $n$.
\item
Assume that  $\alpha_{i+1}$ is a strict prefix of $\alpha_i$ and that $\alpha_i \ul{p_i,\neg Z_i}$ is a prefix of \\ $\alpha_{i+1}\ul{p_{i+1}, Z_{i+1}} \beta_{i+1}$. Thus $\ul{p_i,\neg Z_i}$ occurs in $\beta_{i+1}$, and we can write \\ $\alpha_i=\alpha_{i+1} \ul{p_{i+1}, Z_{i+1}}\delta_1$,  $\,\beta_{i+1}= \delta_1 \ul{p_i,\neg Z_i}\delta_2$ and since $\rho_i$ can be obtained by replacing  $\ul{p_i, \neg Z_i}$ by $\ul{p_i, Z_i} \beta_{i}$ after $\alpha_i$ in $\rho_{i+1}$,
$$\rho_i=\alpha_{i+1}\ul{p_{i+1}, Z_{i+1}}\delta_1\ul{p_i, Z_i} \beta_{i}\delta_2  \gamma_{i+1}$$ follows.
By our assumption on the pair $(\rho_i,\rho_{i+1})$, $\beta_i$ is a terminal path in the body or $Z_{i}$-part of $p_{i}$ and so by
 Proposition \ref{substitute.prop}, $\delta_1\ul{p_{i},  Z_{i}} \beta_{i}\delta_2 $ is a terminal path in the body or $Z_{i+1}$-part of $p_{i+1}$ and so
$\rho_{i+2}$ is obtainable from $\rho_{i}$ by a simple  $l$-reduction, by replacing $\ul{p_{i+1}, \neg Z_{i+1}}\delta_1 \ul{p_{i},  Z_{i}}\beta_{i}\delta_2 $ by $\ul{p_{i+1}, \neg Z_{i+1}}$ in $\rho_{i}$, again contradicting the minimality of $n$.
\item
Lastly, assume that   $\alpha_{i+1}$ is a strict prefix of $\alpha_i$ and that $\alpha_i \ul{p_i,\neg Z_i}$ is not a prefix of  $\alpha_{i+1}\ul{p_{i+1}, Z_{i+1}} \beta_{i+1}$.
  Thus we can write $\alpha_i=
\alpha_{i+1} \ul{p_{i+1}, Z_{i+1}}\beta_{i+1}\, \delta$.  We now  change the order of the two reductions by
 replacing  $\rho_{i+1}$ in the sequence by $\hat{\rho}_{i+1}=\alpha_{i+1}\, \ul{p_{i+1}, \neg Z_{i+1}} \delta\ul{p_{i}, Z_{i}}\beta_i \gamma_i$, which by two applications of
 Proposition \ref{substitute.prop}, is a path through $S$.  In effect we are replacing $ \ul{p_{i+1},  Z_{i+1}}\beta_{i+1}$ by $\ul{p_{i+1}, \neg Z_{i+1}}$ before replacing $ \ul{p_{i},  Z_{i}}\beta_{i}$ by $\ul{p_{i}, \neg Z_{i}}$, instead of in the original order.  Since $\rho_{i+2}=\alpha_{i+1}\, \ul{p_{i+1}, \neg Z_{i+1}} \delta \ul{p_{i},\neg  Z_{i}} \gamma_i$, $\maxpre(\rho_{i},\hat{\rho}_{i+1})$ is  a strict prefix of $\maxpre(\hat{\rho}_{i+1},\rho_{i+2})$.
 Thus, by the minimality of $i$,  after not more than $n-i$ such replacements, the maximal common prefixes of consecutive paths in the  resulting sequence will be strictly increasing in length, as required.
\proofend
\end{enumerate}

\begin{thm}\label{reducible.to.poly.thm}
Let $S$ be a linear schema, let $l$ be a label and  let $\rho,\, \rho' \in \pre(\pathset(S))$. Then it is decidable in polynomial time  whether $\rho$ is $l$-reducible to $\rho'$.
\end{thm}

\proof
By  Lemma \ref{reducible.path.order.lem}, $\rho$ is $l$-reducible to $\rho'$ \ifa\ $\rho$ can be simply $l$-reduced to some $\rho_2\in \pre(\pathset(S))$ \st\ $\rho_2$ is $l$-reducible to $\rho'$ and $\maxpre(\rho,\rho_2)$ is a strict prefix of $\maxpre(\rho_2,\rho')$ and hence $\maxpre(\rho,\rho_2)=\maxpre(\rho,\rho')$. Thus $\rho_2$ exists satisfying these criteria \ifa\ $\rho$ and $\rho'$ have prefixes $\tau\sigma$ and $\tau\sigma'$  respectively \st\ $\sigma'$ is obtained from $\sigma$ by either of the transformations given in Definition \ref{reducible.defn}, and $\rho_2$ is obtained from $\rho$ by replacing $\sigma$ by $\sigma'$. Thus $\sigma$ can be computed in polynomial time if it exists, and this procedure can be iterated using $\rho_2$ in place of $\rho$. The number of iterations needed is bounded by the number of letters in $\rho'$, thus proving the Theorem.
\proofend

\begin{thm}\label{dyn.path.slice.is.a.thm}
Let $S$ be a linear schema containing a label $l$, let $\rho l\in \pre(\pathset(S))$ be executable,   let $V$ be a set of variables and let $S'$ be a quotient of $S$ containing $l$.
\begin{enumerate}
\item
The problem of deciding whether $S'$ is a  $(\rho l,V)$-path-faithful dynamic  slice of $S$ lies in polynomial time.
\item
The problem of deciding whether $S'$ is a  $(\rho l,V)$-dynamic slice of $S$ lies in co-NP.
\end{enumerate}
\end{thm}

\proof
(1) follows immediately from the conclusion of Theorem  \ref{dps.verify.thm}, since given any predicate-free schema $T$ and any variable $v$, the term $\ee{T}{}(v)$ is computable in polynomial time.

To prove (2), we proceed as follows. Any path $\rho'l$ through $S'$ \st\ $\rho'$  is $l$-reducible from $\proj(\rho)$ has length $\le \vert \proj(\rho) l\vert $. We compute a path $\tau$ through $S'$  of length $\le \vert \proj(\rho)l\vert$, with strict inequality \ifa\ $\tau$ is terminal. This can be done in  NP-time by starting with the empty path and successively appending letters to it until a terminal path, or one of length $ \vert \proj(\rho)l\vert$ is obtained. We then test whether $\tau$ is compatible with $\rho$ and does not have a prefix  $\rho'l$ through $S'$ \st\ $\rho'$  is $l$-reducible from $\proj(\rho)$
and  $\ee{\rho}{}(v)=\ee{\rho'}{}(v)$ for all $v\in V$. By Theorem \ref{reducible.to.poly.thm}, this can be done in polynomial time. If  no such  prefix exists for the given $\tau$, then no longer path through $S'$ having prefix $\tau$ has such a prefix either, and hence $S'$ is not  a  $(\rho l,V)$-dynamic  slice of $S$. Conversely, if $S'$ is not  a  $(\rho l,V)$-dynamic slice of $S$, then a path $\tau $ can be computed satisfying the conditions given, proving (2).

\proofend

\begin{thm}\label{dyn.slice.np.comp.thm}
Let $S$ be a linear schema, let $\rho l\in \pre(\pathset(S))$ be executable and  let $V$ be a set of variables.
\begin{enumerate}
\item
The problem of deciding whether \txs\ a non-trivial $(\rho l,V)$-path-faithful dynamic  slice of $S$  is NP-complete.
\item
The problem of deciding whether \txs\ a non-trivial $(\rho l,V)$-dynamic  slice of $S$ lies in PSPACE and  is NP-hard.
\end{enumerate}
\end{thm}

\proof
To prove membership in NP for Problem (1), it suffices to observe that a quotient $S'$ of $S$ can be guessed in NP-time, and using Theorem \ref{dps.verify.thm}, it can be decided in polynomial time whether $S'$ is a non-trivial $(\rho l,V)$-path-faithful dynamic  slice of $S$. Membership of  Problem (2) in PSPACE follows similarly from  Part (2) of Theorem \ref{dyn.path.slice.is.a.thm} and the fact that co-NP$\subb$PSPACE$=$NPSPACE.

To show NP-hardness of both problems, we use a polynomial-time reduction from 3SAT, which is known to be an NP-hard problem \cite{cook:3sat}. An instance of 3SAT comprises a set $\Theta =\{\theta_1,\ldots, \theta_n\}$ and  a propositional formula $\alpha=\bigwedge_{k=1}^{m}\alpha_{k1} \vee \alpha_{k2} \vee \alpha_{k3}$, where each $\alpha_{ij}$ is either $\theta_k$ or $\neg \theta_k$ for some $k$. The problem is satisfied if \txs\ a valuation $\delta: \Theta \to \tf$ under which $\alpha $ evaluates to $\tru$.
 We will construct a linear schema $S$ containing a variable $v$ and a terminal path $\rho$ through $S$ \st\ $S$ has a non-trivial $(\rho,v)$-dynamic  end slice \ifa\ $\alpha $ is satisfiable, in which case this quotient is also a $(\rho,v)$-path-faithful dynamic  end slice.
The schema $S$ is as in Figure \ref{np.hard.fig}.
 \begin{figure}[t]

\begin{center}$ \begin{array}{lll}                  \whi p(v) &\{ \\
                                         &v \as H(v); \\
                                         & \si q_{good}(v)  \then & x \as g_{good}();\\
                                         & \si q_{bad}(v)  \then &  x \as g_{bad}();\\
\\
&\si q_{link}(v) & \then b \as g_{link}(x);\\
&\si q_{reset}(v) &  \then b \as g_{reset}(); \\
&\si Q_{link/reset}(v) & \then v \as F_{link/reset}(b,v);\\
\\
 & \si q_1(v) &\then x \as g_1(b);\\
&\si q_1'(v)& \then x \as g_1'(b);\\
& \vdots \\
& \si q_n(v) &\then x \as g_n(b);\\
&\si q_n'(v) &\then x \as g_n'(b);\\
\\
                                         & \si Q_{test}(v) &    \then \si q_{test}(x) \then v \as F_{test}(v);  \\
                                         &\}
                                         \end{array}$
\end{center}
\caption{} \label{np.hard.fig}
\end{figure}

We say that the function symbol $g_i$ corresponds to $\theta_i$ and $g_i'$ corresponds to $\neg \theta_i$.
The terminal path $\rho$ passes a total of $4 +3n +6n(n-1)  +m $ times through the body of $S$, and then leaves the body.
The paths within the body of $S$ are of fourteen types, and are listed  as follows, in the order in which they occur along $\rho$; note that only those of type (5) depend on the value of $\alpha$.  The total number of paths of each type is given in parentheses at the end.

\begin{enumerate}
\item[($0$)] $ $
\begin{enumerate}
\item[(0.1)]
 $\rho$ passes through $g_{good}$, $g_{link}$, and $ F_{link/reset}$, and through no other   \ass\ apart from $H$.
\item[(0.2)]
$\rho$ passes through  $g_{reset}$, and $ F_{link/reset}$, and through no other   \ass\ apart from $H$.
\item[(0.3)]
$\rho$ passes through $g_{bad}$, $g_{link}$, and $ F_{link/reset}$, and through no other   \ass\ apart from $H$.
\end{enumerate}
(3 paths)
\item
 $\rho$ passes through $g_{good}$ and $F_{test}$, and through no other  \ass\ apart from $H$. (1 path)
\item
 For each $i\le n$,  $\rho$ passes through $g_{good}$, $g_{reset}$, $g_i$ and $F_{test}$ and through no other   \ass\ apart from $H$. ($n$ paths)
\item[($2'$)\!]
 As for type (2), but with $g_i'$ in place of $g_i$. ($n$ paths)
\item
  For each $i\le n$,  $\rho$ passes through $g_{good}$, $g_{link}$, $g_i'$ and $F_{test}$ and through no other   \ass\ apart from $H$. ($n$ paths)
\item
 For each $i \neq j\le n$, $\rho$ passes 3 times consecutively through the body of $S$, as follows;
\begin{enumerate}
\item[(4.1)]
 The first time, it passes through $g_{good}$, $g_{reset}$, and  $g_i$, but not through $q_{test}$ or any other  \ass\ apart from $H$.
\item[(4.2)]
 The 2nd time, it passes through $g_{link}$ and  $g_i'$, but not  through $q_{test}$ or any  other  \ass\ apart from $H$.
 \item[(4.3)]
The 3rd time, it passes through $g_{reset}$ and $g_j$ and $F_{test}$, but through no other \ass\ apart from $H$.
\end{enumerate}
($3n(n-1)$ paths)
\item[$(4.1'$\!\!\!\!]$),(4.2'),(4.3')$ As for types (4.1),(4.2),(4.3),  but with $g_j'$ in place of $g_j$. ($3n(n-1)$ paths)
\item
  For each $i \le m$, $\rho$ passes through $g_{bad}$ and $g_{reset}$, and then through the 3 function symbols corresponding to the implicants $\alpha_{i1}, \alpha_{i2},\alpha_{i3}$, and then through $F_{test}$ and  through no other \ass\ apart from $H$. ($m$ paths)
\end{enumerate}

Before continuing with the proof, we first record the following facts about the terminal path $\rho$.
\begin{enumerate}
\item[(a)]
 $\rho$ passes through all three \asss\ to $v$ and through both \asss\ to $b$.
\item[(b)] All three \asss\ to $v$ in $S$ also reference $v$, and hence if \txs\ a terminal path $\sigma$ through any slice $T$ of $S$  \st\  $\ee{\rho}{}(v)=\ee{\sigma}{}(v)$, then the following hold;
\begin{enumerate}
\item[(b0)]
By (a), $T$ contains $H, \, F_{test}$, $F_{link/reset}$ and hence  $g_{link}$, $g_{reset}$, $g_{good}$ and  $g_{bad}$    because of the type (0)  paths, and thus contains the predicates controlling these function symbols.
\item[(b1)]
 By (a), $\sigma$  passes through all the \asss\ to $v$ in $S$ in the same order as $\rho$ does.
\item[(b2)]
  $\sigma$ and $\rho$ enter the body of $p$ the same number of times, namely the depth of the nesting of $H$ in the term $\ee{\rho}{}(v)$.
\item[(b3)]
For any function symbol $f$ in $S$ assigning to $v$ and for all $k\ge 0$,
  $v$ defines the same term after the $k$th occurrence of $f$ in $\rho$ and  $\sigma$, since this term is the unique subterm of $\ee{\rho}{}(v)$ containing $k$ nested  occurrences of $f$ whose outermost function symbol is $f$.
\item[(b4)]
 For any predicate $q$ in $T$ and for all $k\ge 0$,  $\sigma$ and $\rho$ pass the same way through $q$ at the $k$th occurrence of $q$. For $q \not= q_{test}$, this follows from (b3) applied to $H$ or $F_{link/reset}$. For $q=q_{test}$, it follows from (b1) and (b4) applied to $Q_{test}$.
\item[(b5)] $\proj[T](\rho) =\sigma$. For assume $\sigma' \ul{q,Z} \in \pre(\sigma)$, whereas $\sigma' \ul{q,\neg Z} \in \pre(\proj[T](\rho))$, where $\sigma' \ul{q,Z}$ contains $k$ $q$'s; this contradicts (b4) immediately, and hence $\proj[T](\rho) =\sigma$ follows from Lemma \ref{exe.lem} and the fact that $\proj[T](\rho)$ and  $\sigma$ are both terminal paths through $T$.
\end{enumerate}
\item[(c)] $\rho$ never passes through the predicate terms  $q_{test}(g_{bad}())$ or $q_{test}(g_i'(g_{link}(g_i(g_{reset}()))))$.
\item[(d)] For any prefix  $\rho'$  of $\rho$, the term $\ee{\rho'}{}(v)$ does not contain any $g_i$ or $g_i'$; for these symbols, which do not occur on the type (0) paths, assign to $x$, whereas $ F_{link/reset}$, which does not occur on $\rho$ after the type (0) paths, is the only \ass\ to $v$ referencing a variable other than $v$.
\end{enumerate}
\begin{itemize}
\item
$(\Ra)$.
Let $T$ be a non-trivial $(\rho,v)$-DS of $S$. By (b5),   $T$ is a $(\rho,v)$-PFDS of $S$
and by (b0),  $T$ contains all symbols in $S$ apart possibly from some of those of the form $g_i,g_i'$ and the if predicates $q_i,\,q_i'$ controlling them.  Thus it remains only to show that $\alpha $ is satisfiable.

 We first  show that if $T$ does not contain a symbol $g_j$, then for all $i\not= j$, it cannot contain both $g_i$ and $g_i'$. Consider the  type (4.3) path for the values $i,j$. If $T$ contains $g_i$ and $g_i'$, but not $g_j$, then when $q_{test}$ is reached on path $(4.3)$, the predicate term thus defined, built up over paths $(4.1),(4.2),(4.3)$, is $q_{test}(g_i'(g_{link}(g_i(g_{reset}()))))$, which does not occur along the path $\rho$, contradicting Theorem \ref{dps.verify.thm}. By considering type $(4.3')$ paths the same assertion holds for the symbols $g_j'$. Since $T\not=S$ holds, this implies that $T$ contains at most one element of each set $\{g_i,g_i'\}$.

 We now show that for each $i\le m$, $T$ contains at least one symbol corresponding to an element in $\{\alpha_{i1},\alpha_{i2},\alpha_{i3}\}$. If this is false, then the predicate term $q_{test}(g_{bad}())$, which does not occur along the path $\rho$, would be defined on the $i$th type (5) path, contradicting
 Theorem \ref{dps.verify.thm}.

 Thus $\alpha$ is satisfied by any  valuation $\delta $ \st\ for all $i \le n$,  $\,T$ contains $g_i$ $ \Ra\delta(\theta_i)=\tru$ and  $T$ contains $g_i'$ $ \Ra\delta(\neg \theta_i)=\tru$; since $T$ contains at most one element of each set $\{g_i,g_i'\}$, such a valuation exists.
\item
$(\La)$.
Conversely, suppose that $\alpha$ is satisfiable by a  valuation $\delta: \Theta \to \tf$, and let $T$ be the quotient of $S$ which contains each $g_i$ and $q_i$ \ifa\ $\delta(\theta_i)=\tru$, and containing $g_i'$ and $q_i'$ otherwise, and contains all the other symbols of $S$.
We show that $T$ is  a $(\rho,v)$-DPS of $S$. By Theorem \ref{dps.verify.thm},  it suffices to show that all predicate terms occurring along
$\proj[T](\rho)$ also occur along $\rho$ with the same associated value from $\tf$, since by (d), $\ee{\proj[T](\rho)}{}(v)=\ee{\rho}{}(v)$.

By (d), all predicate terms occurring in $\proj[T](\rho)$  but not $\rho$ must occur  at $q_{test}$ rather than at a predicate referencing $v$. We consider each path type separately and show that no such predicate terms exist.
 \begin{enumerate}
\item[(0)] These paths do not pass through $q_{test}$.
\item $\proj[T](\rho)$ defines $q_{test}(g_{good}())$, which also occurs along $\rho$ in the  type (1) path.
\item If $T$ does not contain $g_i$ then $\proj[T](\rho)$ defines  $q_{test}(g_{good}())$, which occurs along $\rho$ in the type (1) path. Otherwise  $\proj[T](\rho)$ defines $q_{test}(g_i(g_{reset}()))$, which occurs along $\rho$ in a type (2) path.
\item[($2'$)\!] Similar to  type (2).
\item[($3$)] If $T$ does not contain $g_i'$ then $\proj[T](\rho)$ defines  $q_{test}(g_{good}())$, which occurs along $\rho$ in the  type (1) path.  Otherwise  $\proj[T](\rho)$ defines  $q_{test}(g_i'(g_{link}(g_{good}())))$, which also occurs along $\rho$ in a type (3) path.
\item[($4$)]  If $T$  contains $g_i$ but not  $g_j$ or $g_i'$ then $\proj[T](\rho)$ defines  $q_{test}(g_i(g_{reset}()))$, which occurs along $\rho$ in a type (2) path. If $T$ contains $g_i'$   but not $g_j$ or $g_i$ then $\proj[T](\rho)$ defines $q_{test}(g_i'(g_{link}(g_{good}())))$, which also occurs along $\rho$ in a type (3) path. Lastly,
if $T$ contains  $g_j$  then $\proj[T](\rho)$ defines $q_{test}(g_j(g_{reset}()))$, which occurs along $\rho$ in a type (2) path.
\item[($4'$)\!]  Similar to  type (4).
\item[$(5)$] Since  the valuation $\delta$ satisfies $\alpha$, for each $k\le m$,  $T$ contains at least one of
 the 3 function symbols corresponding to the implicants $\alpha_{k1}, \alpha_{k2},\alpha_{k3}$, and hence $\proj[T](\rho)$ defines $q_{test}(g_i(g_{reset}()))$ or  $q_{test}(g_i'(g_{reset}()))$ for some $i \le n$, which occur along $\rho$ in a type (2) or $(2')$ path.
\end{enumerate}
\end{itemize}
Since the schema $S$ and the path $\rho$ can clearly constructed in polynomial time from the formula $\alpha$, this concludes the proof of the Theorem.
\proofend

\section{Conclusion and further directions} \label{conclusion.sect}

We have reformulated Korel and Laski's definition of a dynamic slice of a program as applied to linear schemas, which is the normal level of program abstraction assumed by slicing algorithms, and have also given a less restrictive slicing definition. In addition, we have given P and co-NP complexity bounds for the problem of deciding whether a given quotient of a linear schema satisfies them.  We conjecture  that the problem of whether a quotient $S'$ of a linear schema $S$ is a general dynamic slice \wrt\ a given path and variable set is co-NP-complete. Future work should attempt to resolve this.

 We have also shown that it is not possible to decide in polynomial time whether a given linear schema has a non-trivial dynamic  slice using either definition, assuming  P$\not=$NP. It is possible that this NP-hardness result can be strengthened to PSPACE-hardness for general dynamic  slices, since in this case the  problem does not appear to lie in NP.

We have also shown that minimal   dynamic  slices (whether or not path-faithful) are not unique.
 Placing further  restrictions on either the schemas or the paths may  ensure uniqueness of dynamic  slices or lower the complexity bounds proved in Section \ref{decide.dps.sect},
 and this should be investigated.

Schemas correspond to single programs/methods and so results regarding schemas cannot be directly applied when
analysing a program that has multiple procedures and thus the results in this paper do not apply to inter-procedural slicing.
It would be interesting to extend schemas with procedures and then analyse both dynamic slicing and static
slicing for such schemas.

These results have several practical ramifications. First, since the problem of deciding whether a linear schema has a non-trivial dynamic slice is computationally hard this result must also hold for programs. A further consequence is that the problem of producing minimal dynamic slices must also be computationally hard. We also defined a new notion of a dynamic slice for linear schemas (and so for programs) that places strictly weaker constraints on the slice and so can lead to smaller dynamic slices. Finally, the fact that minimal dynamic slices need not be unique suggests that algorithms that identify and then delete one statement at a time can lead to suboptimal dynamic slices.

\bibliographystyle{elsart-num}
\bibliography{slice240605}

\begin{thebibliography}{10}
\expandafter\ifx\csname url\endcsname\relax
  \def\url#1{\texttt{#1}}\fi
\expandafter\ifx\csname urlprefix\endcsname\relax\def\urlprefix{URL }\fi

\bibitem{greibach:theory}
S.~Greibach, Theory of program structures: schemes, semantics, verification,
  Vol.~36 of Lecture Notes in Computer Science, Springer-Verlag Inc., New York,
  NY, USA, 1975.

\bibitem{delucia:understanding}
A.~{De Lucia}, A.~R. Fasolino, M.~Munro, Understanding function behaviours
  through program slicing, in: $4^{th}$ {IEEE} {W}orkshop on {P}rogram
  {C}omprehension, {IEEE} {C}omputer {S}ociety {P}ress, {L}os {A}lamitos,
  {C}alifornia, {USA}, Berlin, {G}ermany, 1996, pp. 9--18.

\bibitem{mhetal:icsm01}
M.~Harman, R.~M. Hierons, S.~Danicic, J.~Howroyd, C.~Fox, Pre/post conditioned
  slicing, in: {IEEE I}nternational {C}onference on {S}oftware {M}aintenance
  ({ICSM'01}), {IEEE} {C}omputer {S}ociety {P}ress, {L}os {A}lamitos,
  {C}alifornia, {USA}, Florence, Italy, 2001, pp. 138--147.

\bibitem{canfora:salvaging}
G.~Canfora, A.~Cimitile, A.~{D}e {L}ucia, G.~A.~D. {L}ucca, Software salvaging
  based on conditions, in: International {C}onference on {S}oftware
  {M}aintenance (ICSM'96), {IEEE} {C}omputer {S}ociety {P}ress, {L}os
  {A}lamitos, {C}alifornia, {USA}, {V}ictoria, {C}anada, 1994, pp. 424--433.

\bibitem{cimitile:specification}
A.~Cimitile, A.~{D}e {L}ucia, M.~Munro, A specification driven slicing process
  for identifying reusable functions, Software maintenance: Research and
  Practice 8 (1996) 145--178.

\bibitem{gallagher:surgeon}
K.~B. Gallagher, Evaluating the surgeon's assistant: Results of a pilot study,
  in: Proceedings of the International Conference on Software Maintenance,
  {IEEE} {C}omputer {S}ociety {P}ress, {L}os {A}lamitos, {C}alifornia, {USA},
  1992, pp. 236--244.

\bibitem{gallagher:maintenance}
K.~B. Gallagher, J.~R. Lyle, Using program slicing in software maintenance,
  IEEE Transactions on Software Engineering 17~(8) (1991) 751--761.

\bibitem{agrawal:debugging}
H.~Agrawal, R.~A. De{M}illo, E.~H. Spafford, Debugging with dynamic slicing and
  backtracking, Software Practice and Experience 23~(6) (1993) 589--616.

\bibitem{kamkar:thesis}
M.~Kamkar, Interprocedural dynamic slicing with applications to debugging and
  testing, {P}h{D} {T}hesis, {D}epartment of {C}omputer {S}cience and
  {I}nformation {S}cience, {L}ink{\"{o}}ping {U}niversity, {S}weden, available
  as {L}ink{\"{o}}ping {S}tudies in {S}cience and {T}echnology,
  {D}issertations, Number 297 (1993).

\bibitem{lyle:debugging}
J.~R. Lyle, M.~Weiser, Automatic program bug location by program slicing, in:
  $2^{nd}$ International Conference on Computers and Applications, {IEEE}
  {C}omputer {S}ociety {P}ress, {L}os {A}lamitos, {C}alifornia, {USA}, Peking,
  1987, pp. 877--882.

\bibitem{weiser:experiments}
M.~Weiser, J.~R. Lyle, Experiments on slicing--based debugging aids, Empirical
  studies of programmers, Soloway and Iyengar (eds.), Molex, 1985, Ch.~12, pp.
  187--197.

\bibitem{weiser:slicing84}
M.~Weiser, Program slicing, IEEE Transactions on Software Engineering 10~(4)
  (1984) 352--357.

\bibitem{sdetal:lpr}
S.~Danicic, C.~Fox, M.~Harman, R.~Hierons, J.~Howroyd, M.~R. Laurence, Static
  program slicing algorithms are minimal for free liberal program schemas, The
  Computer Journal 48~(6) (2005) 737--748.

\bibitem{laurence:flfl}
M.~R. Laurence, Characterising minimal semantics-preserving slices of
  function-linear, free, liberal program schemas, Journal of Logic and
  Algebraic Programming 72~(2) (2005) 157--172.

\bibitem{korel:dynamic-slicing}
B.~Korel, J.~Laski, Dynamic program slicing, Information Processing Letters
  29~(3) (1988) 155--163.

\bibitem{agrawal:90dynamic}
H.~Agrawal, J.~R. Horgan, Dynamic program slicing, in: Proceedings of the {ACM}
  {SIGPLAN} '90 Conference on Programming Language Design and Implementation,
  Vol.~25, White Plains, NY, 1990, pp. 246--256.
\newline\urlprefix\url{citeseer.ist.psu.edu/agrawal90dynamic.html}

\bibitem{beszedes:dynamic}
A.~Besz\'edes, T.~Gergely, Z.~M. Szab\'o, J.~Csirik, T.~Gyim\'othy, Dynamic
  slicing method for maintenance of large {C} programs, in: Proceedings of the
  Fifth European Conference on Software Maintenance and Reengineering (CSMR
  2001), IEEE Computer Society, 2001, pp. 105--113.

\bibitem{gopal:dynamic-slicing}
R.~Gopal, Dynamic program slicing based on dependence graphs, in: IEEE
  Conference on Software Maintenance, 1991, pp. 191--200.

\bibitem{kamkar:interprocedural:dynamic}
M.~Kamkar, N.~Shahmehri, P.~Fritzson, Interprocedural dynamic slicing, in:
  PLILP, 1992, pp. 370--384.

\bibitem{kamkar:application}
M.~Kamkar, Application of program slicing in algorithmic debugging, in:
  M.~Harman, K.~Gallagher (Eds.), Information and Software Technology Special
  Issue on Program Slicing, Vol.~40, Elsevier, 1998, pp. 637--645.

\bibitem{korel:gotos}
B.~Korel, Computation of dynamic slices for programs with arbitrary control
  flow, in: M.~Ducass{\'e} (Ed.), $2^{nd}$ {I}nternational {W}orkshop on
  {A}utomated {A}lgorithmic {D}ebugging ({AADEBUG'95}), {S}aint--{M}alo,
  {F}rance, 1995.

\bibitem{korel:ist-paper}
B.~Korel, J.~Rilling, Dynamic program slicing methods, in: M.~Harman,
  K.~Gallagher (Eds.), Information and Software Technology Special Issue on
  Program Slicing, Vol.~40, Elsevier, 1998, pp. 647--659.

\bibitem{mueller-olm:precise.parallel}
M.~M�ller-Olm, Precise interprocedural dependence analysis of parallel
  programs, Theoretical Computer Science (TCS) 31~(1) (2004) 325--388.

\bibitem{mueller-olm.seidl:poly.prog.inv}
M.~M�ller-Olm, H.~Seidl, Computing polynomial program invariants, Information
  Processing Letters (IPL) 91~(5) (2004) 233--244.

\bibitem{mueller-olm.seidl:interp.prog.lin.alg}
M.~M�ller-Olm, H.~Seidl, Precise interprocedural analysis through linear
  algebra, in: Proceedings of Principles of Programming Languages (POPL'04),
  Venice, Italy, 2004.

\bibitem{paterson:thesis}
M.~S. Paterson, Equivalence problems in a model of computation, Ph.D. thesis,
  University of Cambridge, {UK} (1967).

\bibitem{luckham:formalised}
D.~C. Luckham, D.~M.~R. Park, M.~S. Paterson, On formalised computer programs,
  J. of Computer and System Sciences 4~(3) (1970) 220--249.

\bibitem{ashcroft:while-goto-siam}
E.~A. Ashcroft, Z.~Manna, Translating program schemas to while-schemas, SIAM
  Journal on Computing 4~(2) (1975) 125--146.

\bibitem{ianov:logical}
Y.~I. Ianov, The logical schemes of algorithms, in: Problems of Cybernetics,
  Vol.~1, Pergamon Press, New York, 1960, pp. 82--140.

\bibitem{rutledge:ianov.schemata}
J.~D. Rutledge, On {Ianov's} program schemata, J. ACM 11~(1) (1964) 1--9.

\bibitem{constable:ianov.schemas}
H.~B. Hunt, R.~L. Constable, S.~Sahni, On the computational complexity of
  program scheme equivalence, SIAM J. Comput 9~(2) (1980) 396--416.

\bibitem{sabelfeld:algorithm}
V.~K. Sabelfeld, An algorithm for deciding functional equivalence in a new
  class of program schemes, Journal of Theoretical Computer Science 71 (1990)
  265--279.

\bibitem{mletal:cfl}
M.~R. Laurence, S.~Danicic, M.~Harman, R.~Hierons, J.~Howroyd, Equivalence of
  conservative, free, linear program schemas is decidable, Theoretical Computer
  Science 290 (2003) 831--862.

\bibitem{lfl:tr}
M.~R. Laurence, S.~Danicic, M.~Harman, R.~Hierons, J.~Howroyd, Equivalence of
  linear, free, liberal, structured program schemas is decidable in polynomial
  time, Tech. Rep. ULCS-04-014, University of Liverpool, electronically
  available at {\tt http://www.csc.liv.ac.uk/research/techreports/} (2004).

\bibitem{mletal:lfl}
S.~Danicic, M.~Harman, R.~Hierons, J.~Howroyd, M.~R. Laurence, Equivalence of
  linear, free, liberal, structured program schemas is decidable in polynomial
  time, Theoretical Computer Science 373~(1-2) (2007) 1--18.

\bibitem{manna:book}
Z.~Manna, Mathematical Theory of Computation, McGraw--Hill, 1974.

\bibitem{BinkleyDGHKK06}
D.~Binkley, S.~Danicic, T.~Gyim{\'o}thy, M.~Harman, {\'A}.~Kiss, B.~Korel,
  Theoretical foundations of dynamic program slicing, Theoretical Computer
  Science 360~(1--3) (2006) 23--41.

\bibitem{HarmanFHHDW02}
M.~Harman, C.~Fox, R.~M. Hierons, L.~Hu, S.~Danicic, J.~Wegener, Vada: A
  transformation-based system for variable dependence analysis, in: SCAM, IEEE
  Computer Society, 2002, pp. 55--64.

\bibitem{jones96}
B.~F. Jones, H.-H. Sthamer, D.~E. Eyres, Automatic structural testing using
  genetic algorithms, The Software Engineering Journal 11~(5) (1996) 299--306.

\bibitem{wegener96}
J.~Wegener, K.~Grimm, M.~Grochtmann, H.~Sthamer, B.~F. Jones, Systematic
  testing of real-time systems, in: 4th International Conference on Software
  Testing Analysis and Review ({EuroSTAR} 96), 1996.

\bibitem{wegener97}
J.~Wegener, H.~Sthamer, B.~F. Jones, D.~E. Eyres, Testing real-time systems
  using genetic algorithms, Software Quality 6~(2) (1997) 127--135.

\bibitem{cook:3sat}
S.~A. Cook, The complexity of theorem-proving procedures, in: STOC '71:
  Proceedings of the third annual ACM symposium on Theory of computing, ACM,
  New York, NY, USA, 1971, pp. 151--158.

\end{thebibliography}

\end{document}